# Interlayer Decoupling in 30 °Twisted Bilayer Graphene Quasicrystal

*Bing Deng[1,†], Binbin Wang[2,†], Ning Li[3,4,†], Rongtan Li[5], Yani Wang[1], Jilin Tang[1,6], Qiang Fu[5], Zhen Tian[2], Peng Gao[3,4]\*, Jiamin Xue[2]\*, and Hailin Peng[1]\**

[1]Center for Nanochemistry (CNC), Beijing Science and Engineering Center for Nanocarbons, Beijing National Laboratory for Molecular Sciences (BNLMS), College of Chemistry and Molecular Engineering, Peking University, Beijing 100871, China

[2]School of Physical Science and Technology, ShanghaiTech University, Shanghai 201210, China

[3]International Center for Quantum Materials, and Electron Microscopy Laboratory, School of Physics, Peking University, Beijing 100871, China

[4]Collaborative Innovation Center of Quantum Matter, Beijing 100871, China

[5]State Key Laboratory of Catalysis, Collaborative Innovation Center of Chemistry for Energy Materials (iChEM), Dalian Institute of Chemical Physics, Chinese Academy of Sciences, Dalian 116023, China

[6]Academy for Advanced Interdisciplinary Studies, Peking University, Beijing 100871, China

[†]These authors contributed equally to this work.

*Correspondence and requests for materials should be addressed to H.P. (hlpeng@pku.edu.cn), or to J.X. (xuejm@shanghaitech.edu.cn), or to P.G (p-gao@pku.edu.cn).

**ABSTRACT**

Stacking order has strong influence on the coupling between the two layers of twisted bilayer graphene (BLG), which in turn determines its physical properties. Here, we report the investigation of the interlayer coupling of the epitaxially grown single-crystal $30°$ twisted BLG on Cu(111) at the atomic scale. The stacking order and morphology of BLG is controlled by a rationally designed two-step growth process, that is, the thermodynamically controlled nucleation and kinetically controlled growth. The crystal structure of the $30°$-twisted bilayer graphene ($30°$-tBLG) is determined to have the quasicrystal like symmetry. The electronic properties and interlayer coupling of the $30°$-tBLG is investigated using scanning tunneling microscopy (STM) and spectroscopy (STS). The energy-dependent local density of states (DOS) with *in-situ* electrostatic doping shows that the electronic states in two graphene layers are decoupled near the Dirac point. A linear dispersion originated from the constituent graphene monolayers is discovered with doubled degeneracy. This study contributes to controlled growth of twist-angle-defined BLG, and provides insights of the electronic properties and interlayer coupling in this intriguing system.

The electronic structure of graphene can be engineered by layer numbers and the stacking orders.[1-3] The twist angle (θ) between the layers of BLG determines the degree of interlayer coupling, and has a crucial role in its electronic properties. AB-stacked bilayer graphene (θ = 0°) exhibits a parabolic band structure[1, 4] with tunable bandgap under vertical electric fields.[2, 5] For the BLG with small twist angles (θ ≤ 5°), the Dirac band structures change dramatically owing to strong interlayer coupling.[6, 7] Specifically, when the twist angle is close to the magic angle (θ ≈ 1.1°),[8] the band structures near Fermi energy become flat,[9] further leading to the unconventional superconductors in the BLG superlattices.[10] Even at relatively large twist angles (θ ≥ 5°), the interlayer coupling between the two monolayers of BLG introduces band hybridization away from the Dirac point,[11] leading to the formation of van Hove singularities (VHSs).[12, 13]

In particular, the 30°-tBLG (θ = 30°) has the largest interlayer twist angle, and is regarded as a quasicrystal with 12-fold rotational symmetry.[14] Recently, the 30°-tBLG has been successfully synthesized on SiC(0001) surface,[15] Pt(111) substrate,[16] and Ni-Cu gradient alloy.[17] Anomalous interlayer coupling with quasi-periodicity[15] was observed with angle resolved photoelectron spectroscopy (ARPES). Multiple replicas of the original 12 Dirac cones appeared and were explained with generalifzed Umklapp scattering.[16] These works point to the possible interlayer interaction in the incommensurate 30°-tBLG, which has previously been shown to be minimal in transport and ARPES measurements of large-angle twisted (although not exactly at 30°) BLG, due to the strong mismatch of crystal momenta in the two layers.[18, 19] We note that the electronic structures of the 30°-tBLG are acquired directly on the growth substrates by ARPES in these pioneer works.[15, 16] This cannot avoid the influence of the substrates, which could have strong interaction with the 30°-tBLG overlayer.

Moreover, the reported growth methods of the 30°-tBLG are complicate and not convenient for transferring it onto other substrates, hindering the study of the intrinsic properties of 30°-tBLG and the applications. Hence, it is highly necessary to grow the 30°-tBLG in an efficient manner, and investigate its intrinsic electronic properties and interlayer coupling by precluding the effect of the substrates.

Here, we report the epitaxial growth of BLG with exclusive 0° and 30° twist angles on Cu(111) by chemical vapor deposition (CVD) and investigate their interlayer coupling at the atomic scale. The stacking order and morphology of the BLG are precisely tailored by the thermodynamics-controlled nucleation process and the kinetics-controlled growth process. The crystal structure of the 30°-tBLG is identified by extensive characterizations including low energy electron diffraction (LEED), scanning transmission electron microscopy (STEM), and selected area electron diffraction (SAED). Then, the 30°-tBLG flakes are transferred onto a flat hexagonal boron nitride (hBN) substrate, and the atomic configuration and electronic structure are investigated by STM and STS. With *in-situ* electrostatic doping induced by the back-gate voltage, we quantitatively show that DOS of the 30°-tBLG can be seen as a sum of two decoupled graphene monolayers and the linear electron dispersion relation retains, indicating the decoupled interlayer interaction near the Dirac point.

**RESULTS AND DISCUSSION**

Cu(111) was chosen as the template for the growth of BLG because of the well-established epitaxial relationship between graphene and Cu(111).[20-22] Here, a two-step process is designed to control the stacking order and morphology of graphene layers with exclusive 0° and 30° orientation related to the copper (Cu) lattice (Figure 1a).

During the nucleation process, the orientation of graphene on Cu(111) is thermodynamically determined by the graphene/Cu step interface.[23-25] The zigzag (ZZ) and armchair (AC) edges of graphene can efficiently saturate the Cu step atoms on Cu(111) and have the largest binding energy (Figure 1a, top inset).[23-26] At the same time with the first layer graphene growth, carbon atoms can diffuse underneath the graphene edge to initiate the nucleation of second layer graphene on Cu(111), with preferred 0° and 30° orientations related to the Cu lattice. During the growth process, the AC edges of graphene grow much faster than the ZZ edges due to less needed carbon radicals to nucleate each atomic row (Figure 1a, bottom inset).[27, 28] According to the kinetic Wulff construction of crystal growth, the faster growing AC edges will disappear, and finally all the single-crystal graphene flakes will grow as regular hexagons terminated by ZZ edges.[27, 28] After several nucleation and growth cycles, trilayer and even multilayer graphene with precisely controlled stacking order and morphology can be obtained (Figure 1a, middle inset).

Experimentally, large-area Cu(111) foils obtained by long-time annealing of polycrystalline Cu foils[20] were used as the growth substrates (Figure S1). The scanning electron microscopy (SEM) image of graphene grown on Cu(111) shows that the all the graphene flakes are with regular hexagonal shapes (Figure S2). Monolayer, bilayer and multilayer regions are clearly visible with shrinking dimensions. With photoelectron emission microscopy (PEEM) imaging (Figure 1b and c), two types of bilayer regions can be identified. The first one has the inner hexagon oriented in the same direction as the outer hexagon (Figure 1b), and the second one has the inner hexagon rotated by ~ 30° (Figure 1c). To identify the microscopic orientation and stacking order of the flakes, micro-LEED (μ-LEED)[29] was used. For the first type, the bilayer region exhibits only a set of 6 diffraction spots that is identical to the monolayer region (Figure 1b), confirming the AB stacking order. In contrast, for the second type, the diffraction pattern of the bilayer region has 12 spots,[15] which can be regarded as a combination of two sets of 6-fold patterns that is rotated by 30 ° (Figure 1c). The up and bottom layers of the 30 °-tBLG can be distinguished from the μ-LEED pattern because the intensity of the upper layer (blue circles in Figure 1c) is larger than the bottom layer (pink circles in Figure 1c). The μ-LEED results also prove that the smaller graphene layer is beneath the larger layer. Apart from the BLG, trilayer graphene (TLG) were also characterized with exclusive 0 °and 30 °orientations (Figure S3).

The layer number and stacking order of over 1000 graphene domains are statistically shown in Figure 1d and Table S1. For the monolayer graphene (MLG), ~ 98% of them align with the Cu(111) lattice with 0 ° rotation; for the BLG, the AB stacking and 30 °twisting have almost the same probability; for the TLG, the stacking orders can be 0 °-0 °-0 °, 0 °-0 °-30 °, 0 °-30 °-0 °, and 0 °-30 °-30 °, with the probability near 2:1:1:2. One of the merits of graphene grown on Cu foils lies in the feasibility of isolating it from the growth substrates, benefiting the investigation of the intrinsic properties by precluding the effect of substrates. In Figure 1e, we showed that the graphene islands can be transferred onto $SiO_2$/Si substrates without damaging or folding. Due to the good optical contrast of graphene on $SiO_2$/Si substrates, the alignment of graphene layers can be easily identified with an optical microscope (OM), without the complication with advanced microscopy techniques such as PEEM.

To further reveal the crystal structure, the BLG samples were transferred onto a TEM grid by a polymer-free transfer method[30] for TEM and SAED investigation. Here, high angle annual dark field STEM (HAADF-STEM) was carried out thanks to its excellent contrast and spatial resolution in the characterization of graphene.[31, 32] STEM was operated at 60 kV to avoid the damage of graphene by electron beam.[33]

Figure 2a and Figure 2b shows the HAADF-STEM image and SAED pattern of AB-stacked BLG, respectively. The AB-stacked BLG does not exhibits Moiré patterns (Figure 2a), and the enlarged contrast image matches well with the atomic model of the AB stacking order (Figure 2a, inset). In contrast, the STEM image of the 30°-tBLG shows special Moiré pattern without spatial periodicity (Figure 2c), and fits well with the atomic configuration of two overlapped graphene lattices rotated exact 30° (Figure 2c, inset). The SAED of 30°-tBLG exhibits the dodecagonal symmetry (Figure 2d), similar to the LEED results (Figure 1c), but with equal strength of scattering points. The twisted angle of tBLG domains was statistically measured to be $(30.00 \pm 0.41)°$ (Figure S4). Note that the small uncertainty may originate from the mechanical error and/or the local roughness of freestanding graphene on the TEM grid. The STEM image of the 30°-tBLG can be viewed as a dodecagonal quasicrystal lattice, which is entirely filled out by rhombuses, equilateral triangles, and squares with different orientations (Figure 2e, up). All three structural units of the quasicrystal lattice can be identified according to the false-colored STEM images (Figure 2e, below).

Apart from the crystal structure identification, the electron energy loss spectroscopy (EELS) in aberration-corrected STEM (AC-STEM) is an efficient tool to probe the electronic structures of graphene.[34,35] As shown in Figure 2f, compared with the AB-stacked BLG, an additional energy loss peak appears in the vicinity of π dispersion for the 30°-tBLG. This low energy excitation derives from VHSs in the DOS induced by the interlayer coupling in the 30 °-tBLG, [12, 13, 34] which is consistent with the optical absorption in twisted BLG.[26, 35]

The 30 °-BLG is an energetically unfavorable incommensurate structure, and is stabilized by the crystal lattice of the Cu(111) substrate during growth. Hence, it is necessary to clarify its thermal stability without substrate restriction. Here, the *in-situ* heating experiments of the 30 °-BLG freestanding on a TEM grid was carried out (Figure S5). Surprisingly, the 30 °-BLG kept the same diffraction pattern with temperature ramping up to 700 °C, indicating that the quasicrystal 30 °-BLG is stable even without the substrate restriction.

ARPES has been used to study the global band structure of 30 °-tBLG on the growth substrates of SiC and Pt.[15, 16] Here, STM was carried out to study the electronic properties at the atomic level. hBN has proven to provide an atomically flat and clean substrate, which helps to reveal the intrinsic properties of graphene with minimal disturbance of the substrates.[36] 30 °-tBLG was transferred to hBN flakes exfoliated on a silicon (Si) wafer (Figure 3a). Both the bilayer and monolayer regions of the BLG flake have overlaps with the hBN. Then, electron beam lithography (EBL) and metal deposition were used to fabricate the device for STM study (Figure 3b). Metal electrode surrounding the 30 °-tBLG area facilitated the sample locating in the STM chamber and protected the tip from crash. With the highly doped Si as a back gate, the Fermi level in the 30 °-tBLG can be electrostatically tuned, which is crucial for a quantitative determination of the DOS at different energies and the coupling between the two layers, as described later.

After the device was fabricated, atomic force microscope (AFM) was used to check the surface of the 30 °-tBLG on hBN (Figure 3c). Although there were some bubbles and wrinkles on the surface, many large flat areas were available for the STM measurement. Some monolayer area was also exposed for a comparison with the 30°-tBLG (marked with white dashed line in Figure 3c).

The STM study was performed at 4.5 K under ultrahigh vacuum, and high quality scans can be obtained thanks to the ultraflat hBN substrate. The atomically resolved topography of the 30°-tBLG is uniform in large-area (Figure 3d), and shows a clear triangular lattice, as indicated by the zoom-in scan (Figure 3d, inset). Although the real-space STM topographies appear to be similar to those of normal AB-stacked BLG,[37] fast Fourier transform (FFT) reveals the hidden structures (Figure 3e), which is drastically different from those of AB-stacked BLG.[38] The 12 spots in the outer circle can be assigned to the graphene lattice, with the measured lattice constant of ~ 0.24 nm.[37] However, unlike the TEM diffraction of the 30°-tBLG with equal scattering strength (Figure 2d), the 12 spots derived from lattice scattering in the STM FFT show alternating brightness, similar to the μ-LEED pattern in Figure 1c. The dimmer set is rotated by 30° related to the brighter set. We assign the dimmer set to the bottom layer graphene and the brighter set to the top layer graphene. Since the signal of STM mainly comes from the surface of the sample, the top layer graphene provides stronger signals than the bottom layer. In contrast, in the TEM scattering the electrons penetrate the whole stack of the 30°-tBLG and have equal strength of interaction with both layers. This difference indicates that the 30°-tBLG does not possess a strict 12-fold rotational symmetry since the two layers are not in the same plane.

The inner circle of 12 spots in Figure 3e originates from the Moiré pattern of the 30°-tBLG. Each spot can be constructed by the difference of two neighboring reciprocal lattice vectors and one example is shown in Figure 3e. Since each Moiré spot involves two lattice vectors from both the top and bottom layers, it has a very uniform intensity and a true 12-fold rotational symmetry. The Moiré pattern period $L$ follows the function $L = a/2\sin(\theta/2) \approx 0.475$ nm, where $\theta$ is the twisted angle, and $a$ is the lattice constant of graphene. Since the top layer lattice points dominate the topography image of Figure 3d, these Moiré features are not easily observable. However, if we filter out the atomic lattice points and only use the 12 Moiré pattern points to perform the inverse Fourier transform, a topography image closely resembles that of the STEM image (Figure 2c) and a dodecagonal quasicrystal is obtained (Figure 3f).

With the understanding of the topography features, we turn to probe the DOS and interlayer coupling in 30°-tBLG using STS. Figure 4a shows a typical d$I$/d$V$ spectroscopy curve of 30°-tBLG. Similar to the spectroscopy data of AB-stacked BLG,[39] the curve is V shaped with the minimum indicating the position of the Dirac point. The Dirac point locates at about −0.06 eV, indicating that the sample is slightly electron doped. However, due to complicated dependence of tunneling probability on bias voltage,[40] no quantitative information about the DOS can be readily obtained from a single d$I$/d$V$ curve.

Taking the advantage of precise electrostatic doping induced by the Si back gate, we can *in-situ* tune the Fermi level of 30°-tBLG and track the movement of the Dirac point, which gives quantitative information of the DOS as a function of energy. In Figure 4b, the $|(\mathrm{d}^2 I)/(\mathrm{d}V^2)|$ *versus* back gate voltage $V_g$ and bias voltage is plotted. An extra numerical differentiation of the d$I$/d$V$ data is taken to enhance the visibility of the Dirac point, which is fitted with the white line (Figure S6). When $V_g$ is swept from negative to positive, the 30 °-tBLG becomes electron doped and the Dirac point shifts to negative bias voltage as expected. In the case of AB-stacked BLG,[39] the Dirac point shifts linearly with $V_g$ due to its parabolic dispersion relation. Here for the 30 °-tBLG, it is evident that the Dirac point shifts nonlinearly with $V_g$, indicating the very different electronic structures of different stacking orders.

The energy of Dirac points $E_D$ at different $V_g$ was extracted and plotted in Figure 4c. Assuming a linear dispersion in the 30 °-tBLG, the data can be nicely fitted with the following function,[41]

$$E_D = \frac{\hbar \nu_F}{e} \sqrt{\frac{4\pi}{g} \frac{C}{e} (V_g - V_0)} \qquad (1)$$

where $\hbar$ is the Plank constant, $e$ is the electron charge, $\nu_F$ is the Fermi velocity, $C$ is the capacitance between back gate and sample, $V_0$ is back gate voltage at which the Dirac point is at the Fermi level, and $g$ is the degeneracy of 30 °-tBLG. Slightly different Fermi velocities at the electron side and hole side were obtained to be $0.92 \times 10^6$ m/s and $0.88 \times 10^6$ m/s, respectively, reflecting the moderate electron-hole asymmetry.[42, 43] These values are very close to that of monolayer graphene, as previously reported.[41, 44] Interestingly, the degeneracy factor $g$ in the 30 °-tBLG is determined to be 8, while $g$ is 4 in monolayer graphene and accounts for the spin and valley degeneracy.[45] This extra double degeneracy can be traced back to the two layers. If the 30 °-tBLG does not modify the band structure of each constituent monolayer graphene, the Fermi velocities will remain the same and only the DOS doubles. These results clearly reveal the interlayer decoupling of electronic states in the 30 °-tBLG within the energy window probed here. These findings are consistent with the ARPES observations,[15] where the dispersion relation of 30 °-tBLG was found to be the same as that of monolayer graphene. The atom density doubles from monolayer graphene to 30°-tBLG, so the local DOS is increased by a factor of 2.

Due to the finite density of states and insufficient screening in graphene, any external electric field will cause some displacement field between the layers and break the layer degeneracy. The fitting function described by Eq. (1) does not take this effect into account. To deal with it rigorously, we adopted a model developed by Kim *et al*.[46] and get a self-consistent expression between $V_g$ and $E_D$ (see Supplementary Information, Figure S7-8). The new fitting procedure gives results very similar to that of Eq. (1) with slightly different Fermi velocities, indicating that the 30°-tBLG can indeed be treated as two decoupled monolayer graphene.

To verify that the electronic properties of the constituent monolayer graphene are not modified by the growth, device fabrication process and sample configuration, we performed similar spectroscopy study on the same sample where monolayer graphene is exposed. The topography of monolayer graphene on hBN is shown in Figure S9, and the gate-tuned STS results are summarized in Figs. 4d to 4f. The single d$I$/d$V$ curve (Figure 4d) is typical of monolayer graphene on hBN, where the dip can be seen in the hole side due to the formation of second generation Dirac point.[44] We note that the 30°-tBLG does not show any dip feature in the energy window between –0.4 V to +0.4 V (Figure S10). Moreover, the Fermi level is more effectively tuned by the back gate due to reduced DOS (Figure 4e). The fitted Fermi velocities for monolayer graphene are $1.05 \times 10^6$ m/s for holes and $0.85 \times 10^6$ m/s for electrons with a degeneracy of 4 (Figure 4f), consistent with previous reports.[44] These results corroborate the conclusion we made about the interlayer decoupling in 30°-tBLG.

To compare the DOS of 30°-tBLG and AB stacked bilayer graphene in our experiment, we also fabricated devices with AB-stacked bilayer graphene grown on the same copper foil of the 30°-tBLG. Indeed, the Dirac point shift linearly with the back gate voltage (Figure 4g-i) and an effective mass of 0.034 free-electron mass is obtained. This comparison directly reflects the important effects of stacking orders on the electronic structures.

Previous transport study of randomly twisted bilayer graphene indicated that the quantum Hall voltage in this system could be regarded as a combination of two monolayer graphene.[47] ARPES measurements of exactly 30°-tBLG indicated the strong scattering of the electron waves between the layers.[15,16] Our results provide the direct local DOS measurement and especially their response to external gating, which add to the understanding of the electronic structure in 30°-tBLG.

**CONCLUSIONS**

We have successfully synthesized high quality 30°-tBLG on Cu(111) by a rationally designed two-step process, and identified its crystal structure with LEED and TEM. The 30°-tBLG was transferred onto flat hBN substrates, and STM and STS were carried out to study its topography and electronic properties at the atomic level. The interlayer electronic state of the 30°-tBLG was found to be decoupled, as indicated by the measurement of the energy-dependent local DOS with *in-situ* electrostatic doping. A linear dispersion originated from the constituent graphene monolayers is discovered with doubled degeneracy originating from the two layers. Recent theoretical calculation predicted modified band structure at higher energies (> 1 eV).[48] Although these higher-energy features may be important for some optical processes, most of the physical properties in this system such as transport phenomena and so on depend mainly on the electronic dispersion near the Dirac point, which is probed in this study. Our work not only proposed a facile route for controlled growth of 30 ° twisted bilayer graphene, but also provided insights of the electronic properties and interlayer coupling in this intriguing system.

**EXPERIMENTAL SECTIONS**

**Growth of 30°-tBLG on Cu(111).** The Cu(111) foil was obtained by long-time annealing of polycrystalline Cu foil (Alfa Aser, 25 μm, 99.8%, #46365) by using a homemade CVD system. For the electrochemical polishing of the Cu(111) foil, a mixture of phosphoric acid and ethylene glycol ($H_3PO_4$:EtOH, v:v = 3:1) was used as the electrolyte, and the voltage and polishing time were set to 2 V and 30 min, respectively. Graphene growth was conducted using a low pressure CVD (LPCVD) system. The Cu(111) foil was heated to 1030 °C under the gas flow of 500 sccm $H_2$ at the pressure of ~ 500 Pa. After annealing for 30 min under the same gas condition, methane ($CH_4$) with gas flow of 1 sccm was injected into the CVD system. The growth time was usually 20 − 30 min. After growth, the Cu(111) was rapidly cooled down to room temperature.

**Graphene transfer.** For the transfer of graphene onto $SiO_2$/Si substrate, the poly-(methyl methacrylate) (PMMA)-mediated wet etching transfer was applied. A PMMA film was spin-coated on graphene/Cu(111) at the speed of 2000 rpm for 1 min. Then, the PMMA/graphene film was detached from Cu by wet etching Cu in a 0.1 M ammonium persulfate (($NH_4)_2S_2O_6$) (aq). After detaching, the PMMA/graphene was rinsing in DI water for 3 times and attached onto $SiO_2$/Si substrate. After drying in air, the PMMA was dissolved by hot acetone. For the transfer of graphene onto gold (Au) grids for TEM characterization, the polymer-free direct transfer method was used to avoid contamination.[30] For the transfer of graphene onto hBN flakes for STM measurement, the Polydimethylsiloxane (PDMS)-mediated dry transfer process was applied[49] to avoid the water encapsulation onto the graphene–hBN interface and the contamination of graphene surface. The hBN flakes was firstly placed onto the $SiO_2$/Si substrate by mechanical exfoliation. The PDMS stamp was attached to the graphene/Cu(111). Then, the Cu substrate was etched away using $(NH_4)_2S_2O_6$(aq), leaving the graphene adhered on PDMS stamp. The PDMS/graphene was rinsed in DI water for 3 times and totally dried in a vacuum tank. Then, the PDMS/graphene was attached to the hBN on $SiO_2$/Si and the PDMS was manually peeled off, leaving the graphene on the hBN flakes.

**Graphene characterization.** The OM was conducted on a Nikon Olympus LV100ND system. The X-ray diffraction (XRD) was carried out using an X-ray powder diffractometer (Rigaku D/MAX-PC 2500). The SEM images were obtained on a Hitachi S4800 field-emission scanning electron microscope. Electron backscattering diffraction (EBSD) measurements were carried out on ULVAC-PHI (PHI 710) Auger system equipped with the EBSD probe (EDAX, DigView). EBSD test was operated at voltage of 10 kV and current of 10 nA. The spot size was 20 nm and the angular resolution was of the order of 0.1 °. AFM image was carried out on a Bruker Dimension Icon with a Nanoscope V controller using the ScanAsyst mode. PEEM and μ-LEED were conducted in an Elmitec LEEM-III system (base pressure < $10^{-10}$ Torr). The selected area for LEED is about 2 μm and the electron energy was fixed to 50 eV.

**TEM and STEM.** The HAADF-STEM images were obtained by using the Nion U-HERMS200 electron microscope operated at 60 kV, with a 35-mrad convergence half-angle (α/2). The spatial resolution is better than 1.07 Å estimated by the FFT image. The images were processed by a free HREM Digital Micrography plug-in, and the inset images in Figure 2a and Figure 2b were additionally processed by Gaussian blur method. False color was made from Swift powered by Nion Co. EELS data was acquired at 30-mrad convergence half-angle (α/2) and 24.5-mrad collector half-angle (β/2), and the acquisition time of each spectrum is 100 s.

**STM and STS.** The STM and STS measurements were performed under ultrahigh vacuum (pressure $\leq 10^{-11}$ mbar) and liquid-helium temperature with an Omicron low-temperature STM (LT-STM). The tips used were electrochemically etched tungsten wires. The topography images were taken in a constant-current scanning mode. The d$I$/d$V$ spectroscopy was acquired by turning off the feedback loop. A small a.c. modulation of 10 mV at 991.7 Hz was applied to the bias voltage and the corresponding change in current was measured using lock-in detection.

**ASSOCIATED CONTENT**

**Supporting Information**

The supporting information is available free of charge on the ACS Publication website at DOI: XXX.

Figures showing additional characterization of Cu(111), SEM and LEED characterization of graphene, thermal stability of quasicrystal bilayer graphene by *in-situ* TEM heating, gate-voltage dependent STS and the Dirac-point fitting procedure of quasicrystal bilayer graphene, discussion on the effect of displacement field on the band structures, and STM topography and STS spectrum of bilayer graphene (PDF)

**Financial Interest**

The authors declare no competing financial interest.

**AUTHOR INFORMATION**

**Corresponding Author**

*Email (P. Gao): p-gao@pku.edu.cn

*Email (J. Xue): xuejm@shanghaitech.edu.cn

*Email (H. Peng): hlpeng@pku.edu.cn

**ORCID**

Bing Deng: 0000-0003-0530-8410

Qiang Fu: 0000-0001-5316-6758

Peng Gao: 0000-0003-0860-5525

Jiamin Xue: 0000-0002-1892-1743

Hailin Peng: 0000-0003-1569-0238

**Author Contributions**

[†]B. Deng, B. Wang, and N. Li contributed equally to this work. B.D., B.W., J.X., and H.P. conceived the experiment. B.D. conducted the synthesis, transfer and characterization of graphene with the help of J.T. and Y.W. The PEEM and LEED test was carried out by R.L. and Q.F. N.L. performed TEM and STEM experiments and analyzed the data under direction of P.G. The STM was conducted by B.W., Z.T. and J.X. B.D., B.W., J.X., and H.P. wrote the paper. All the authors discussed the results and commented on the manuscript.

**ACKNOWLEDGMENT**

We thank D. Pei from Oxford University and P. Moon from New York University for helpful discussions. This work was financially supported by the Ministry of Science and Technology of China (Grant Nos. 2016YFA0200101 and 2017YFA0305400), the National Natural Science Foundation of China (Grant Nos. 21525310, 51432002, and 51520105003), and Beijing Municipal Science & Technology Commission (Grant No. Z181100004818001).

Oxford University Press: New York, 2008.

**FIGURES**

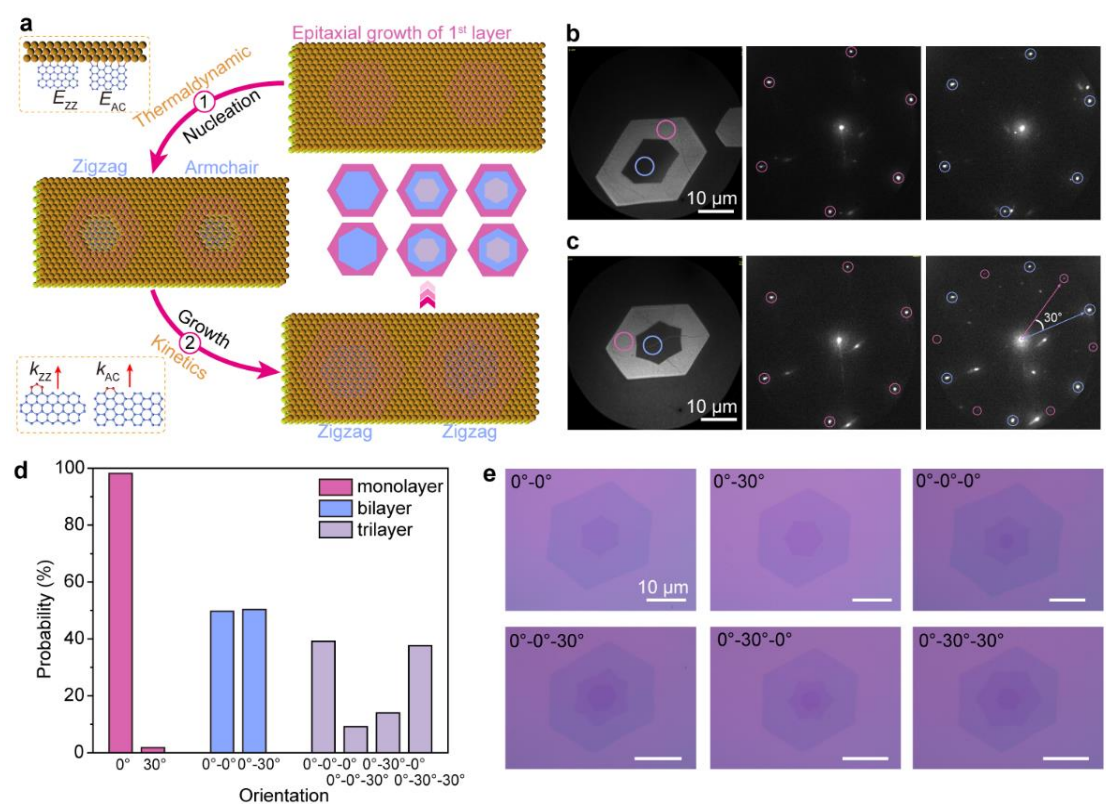

**Figure 1. Epitaxial Growth of 30°-tBLG quasicrystal on Cu(111). (a) Experimental design of the growth of bilayer and multilayer graphene with exclusive 0° and 30° orientation on Cu(111). (b) PEEM image and μ-LEED**

patterns of the AB-stacked BLG. **(c) PEEM image and μ-LEED patterns of the 30°-tBLG. (d) Histogram of the graphene layer number and orientation related to the Cu substrates. The sequence is ordered from the upper layer to the lower layer. (e) OM images of BLG and TLG with various stacking orders.**

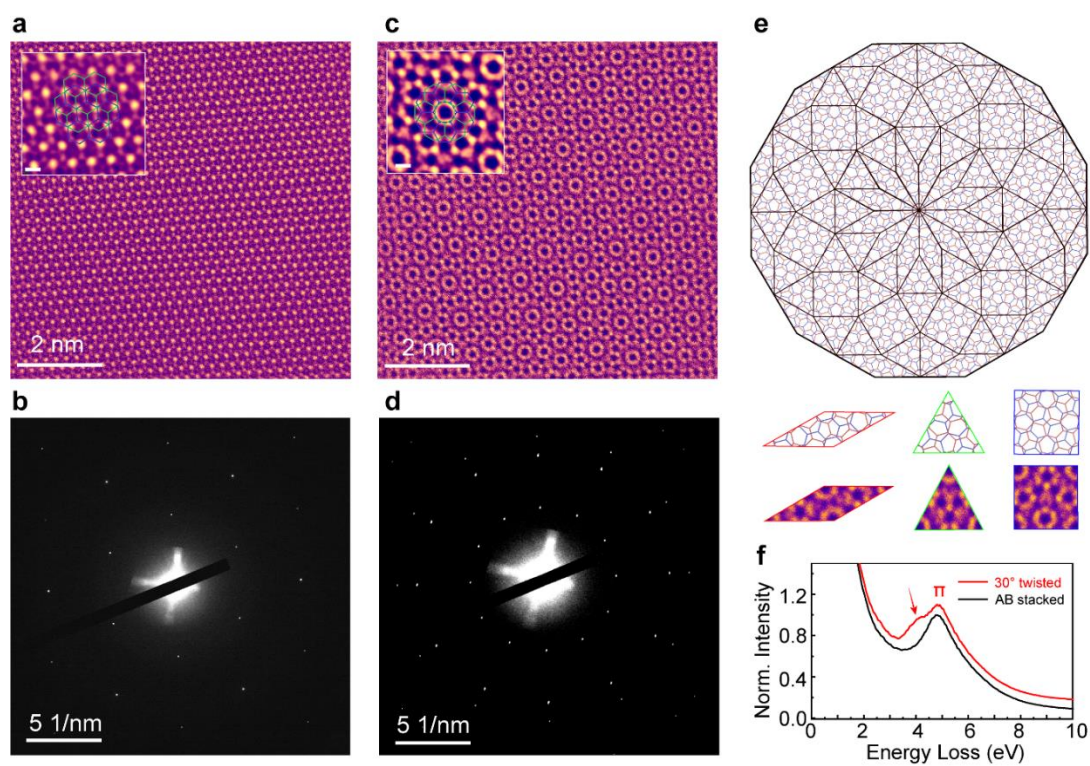

**Figure 2. Crystal structure identification of the 30°-tBLG. (a) HAADF-STEM image of AB-stacked BLG. Inset, enlarged STEM image and the crystal scheme. (b) SAED of AB-stacked BLG. (c) HAADF-STEM image of 30°-tBLG. Inset, enlarged STEM image and the crystal scheme. (d) SAED of 30°-tBLG. (e) Scheme of 30°-tBLG as a dodecagonal quasicrystal pattern, and the structure units. (f) EELS of AB-stacked BLG and 30°-tBLG. Scale bars in insets of a and c are 0.2 nm.**

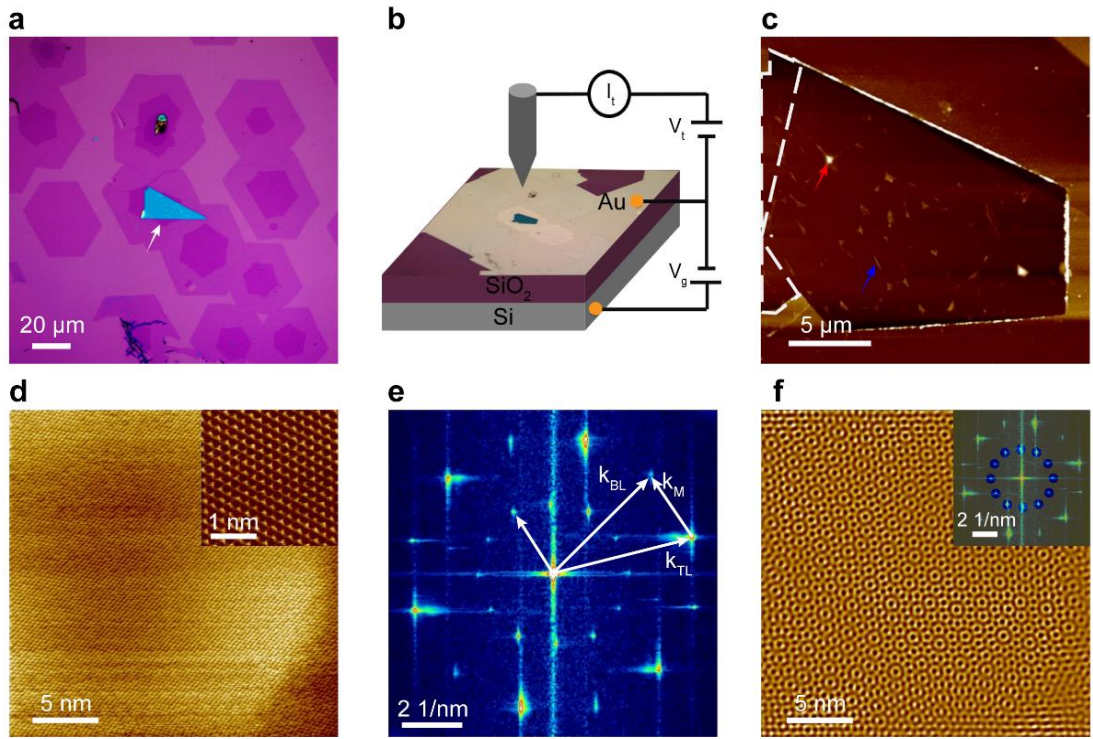

**Figure 3. STM topography of the 30°-tBLG. (a)** OM image of graphene flakes on hBN/SiO$_2$. White arrow indicates the hBN flake. **(b)** Schematic of the STM measurement set up. **(c)** AFM image of the BLG surrounded by the metal electrodes. The area enclosed by the dashed line is the monolayer region, while the rest is the 30°-tBLG region. The red arrow and blue arrow indicate the bubbles and wrinkles present on the BLG, respectively. **(d)** Atomically resolved large-area topography of 30°-tBLG. Inset, zoom in to show the lattice. **(e)** FFT of the topography image in d. $K_{BL}$ and $K_{TL}$ correspond to one of the 6 lattice vectors of the bottom layer and top layer, respectively. $K_M$ corresponds to one of the 12 Moiré vectors. **(f)** Reverse FFT with only the Moiré pattern, showing the quasicrystal like structure. Inset, the 12 points used to conduct the inverse FFT are marked with blue circles.

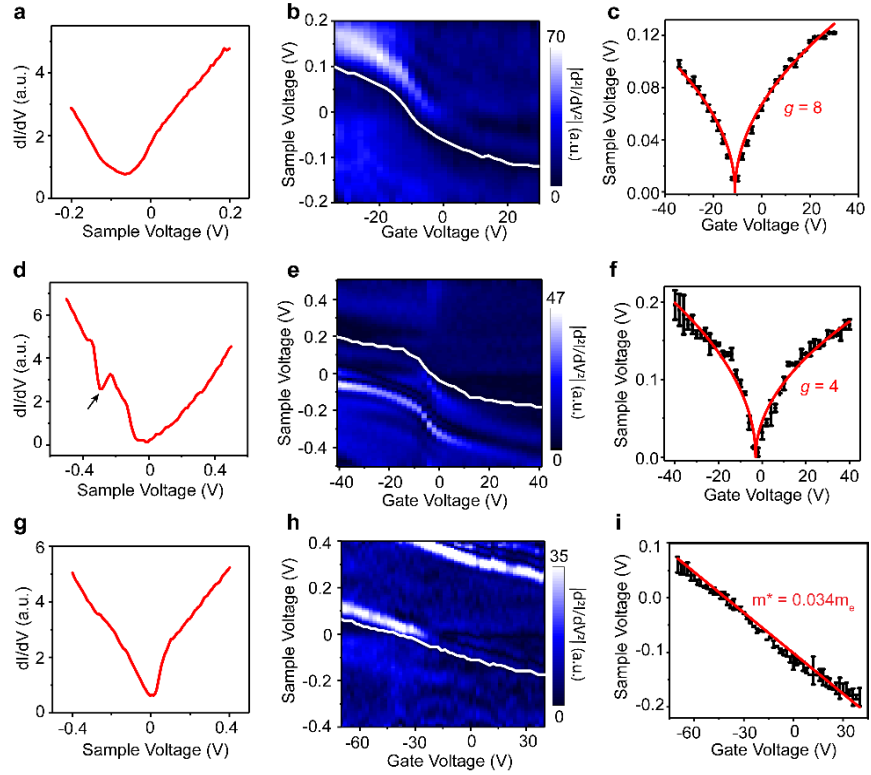

**Figure 4. STS study of 30°-tBLG, MLG, and AB-stacked BLG. (a) Single spot STS of 30°-tBLG at zero gate voltage. (b) STS of 30°-tBLG as a function of gate voltage. (c) Variation of Dirac point energy as a function of gate voltage. The dots are fitted assuming linear dispersion relation and degeneracy of 8. (d) Single spot STS of monolayer graphene at zero gate voltage. The dip at the negative side originates from the secondary Dirac points. (e) STS of monolayer graphene as a function of gate voltage. (f) Variation of Dirac point energy as a function of gate voltage. The dots are fitted with linear dispersion relation and degeneracy of 4. (g) Single spot STS of AB-stacked BLG at -40 V gate voltage. (h) STS of AB-stacked BLG as a function of gate voltage. (i) Variation of Dirac point energy as a function of gate voltage. The dots are fitted with parabolic dispersion relation and the effective mass is $0.034 m_e$. The white line consists of the positions of fitted Dirac points at each gate voltage for b, e, and h.**